\documentclass[12pt]{article}
\usepackage{epsfig}

\newskip\humongous \humongous=0pt plus 1000pt minus 1000pt

\newif\ifdtup

\def\pr#1{#1^\prime}

\def\beq{\begin{equation}}
\def\eeq{\end{equation}}

\def\beqn{\begin{eqnarray}}
\def\eeqn{\end{eqnarray}}
\relax

\def\s{\sigma}
\def\fix{\right|_{\rm FO}\!\!\!\!}
\def\res{\right|_{\rm res}\!\!\!\!}
\def\imp{\right|_{\rm imp}\!\!\!\!}
\def\TNLL{\right|_{\rm TNLL}\!\!}
\def\NLL{\right|_{\rm NLL}\!\!\!\!}
\def\LL{\right|_{\rm LL}\!\!}

\def\lq{\left[}
\def\rq{\right]}
\def\rg{\right\}}
\def\lg{\left\{}
\def\({\left(}
\def\){\right)}
\def\xe{x_E}
\def\xp{x_p}
\def\zp{z_p}

\def\sh{\hat{\sigma}}
\def\qb{\bar{q}}
\def\ah{\hat{a}}
\def\L{\log\frac{\mu^2}{\mu_0^2}}

\newcommand\LambdaQCD{\Lambda_{\scriptscriptstyle \rm QCD}}

\catcode`\@=11

\@addtoreset{equation}{section}
\def\theequation{\thesection.\arabic{equation}}

\def\@normalsize{\@setsize\normalsize{15pt}\xiipt\@xiipt
\abovedisplayskip 14pt plus3pt minus3pt%
\belowdisplayskip \abovedisplayskip
\abovedisplayshortskip \z@ plus3pt%
\belowdisplayshortskip 7pt plus3.5pt minus0pt}

\def\small{\@setsize\small{13.6pt}\xipt\@xipt
\abovedisplayskip 13pt plus3pt minus3pt%
\belowdisplayskip \abovedisplayskip
\abovedisplayshortskip \z@ plus3pt%
\belowdisplayshortskip 7pt plus3.5pt minus0pt
\def\@listi{\parsep 4.5pt plus 2pt minus 1pt
     \itemsep \parsep
     \topsep 9pt plus 3pt minus 3pt}}

\@twosidetrue

\relax

\catcode`@=12

\catcode`\@=11

\def\section{\@startsection{section}{1}{\z@}{3.5ex plus 1ex minus
   .2ex}{2.3ex plus .2ex}{\large\bf}}

\def\thesection{\arabic{section}}

\def\appendix{\setcounter{section}{0}
 \def\thesection{\Alph{section}}
 \def\theequation{\Alph{section}.\arabic{equation}}
 \def\section{\@startsection{section}{1}{\z@}{3.5ex plus 1ex minus
   .2ex}{2.3ex plus .2ex}{\large\bf}}
 \def\subsection{\@startsection{section}{2}{\z@}{3.25ex plus 1ex minus
   .2ex}{1.5ex plus .2ex}{\large\bf}}}

\def\ps@headings{\def\@oddfoot{}\def\@evenfoot{}
\def\@oddhead{\hbox{}\hfill
 \makebox[.5\textwidth]{\raggedright\ignorespaces --\thepage{}--
 \hfill {}}}  
\def\@evenhead{\@oddhead}
\def\subsectionmark##1{\markboth{##1}{}}
}

\ps@headings

\catcode`\@=12

\def\figcap{\section*{Figure Captions\markboth
 {FIGURECAPTIONS}{FIGURECAPTIONS}}\list
 {Fig. \arabic{enumi}:\hfill}{\settowidth\labelwidth{Fig. 999:}
 \leftmargin\labelwidth
 \advance\leftmargin\labelsep\usecounter{enumi}}}
 \relax
\def\tablecap{\section*{Table Captions\markboth
 {TABLECAPTIONS}{TABLECAPTIONS}}\list
 {Table \arabic{enumi}:\hfill}{\settowidth\labelwidth{Table 999:}
 \leftmargin\labelwidth
 \advance\leftmargin\labelsep\usecounter{enumi}}}
 \relax
\def\reflist{\section*{References\markboth
 {REFLIST}{REFLIST}}\list
 {[\arabic{enumi}]\hfill}{\settowidth\labelwidth{[999]}
 \leftmargin\labelwidth
 \advance\leftmargin\labelsep\usecounter{enumi}}}
 \relax

\catcode`\@=11

\relax

\catcode `@ 11
\def\biblabel#1{\if@filesw\immediate
\write\@auxout{\string\bibcite{#1}{\the\value{\@listctr }}}\fi}
\catcode `@ 12

\newcommand{\ccaption}[2]{
  \begin{center}
    \parbox{0.85\textwidth}{
      \caption[#1]{\small\it {#2}}}
  \end{center}    }
\def    \be             {\begin{equation}}
\def    \ee             {\end{equation}}
\def    \ba             {\begin{eqnarray}}
\def    \ea             {\end{eqnarray}}
\def    \nn             {\nonumber}
\def    \=              {\;=\;}
\def    \frac           #1#2{{#1 \over #2}}

\def \ep{\epsilon}
\def \as   {\ifmmode \alpha_s \else $\alpha_s$ \fi}

\def\b0{b_0}

\def \mt   {\ifmmode m_{\rm t} \else $m_{\rm t}$ \fi}

\def \to   {\mbox{$\rightarrow$}}
\newcommand     \MSB            {\ifmmode {\overline{\rm MS}} \else
                                 $\overline{\rm MS}$\fi}

\newcommand\hepph[1]{{\tt hep-ph/#1}}

\newcommand     \epem           {\ifmmode{e^+e^-}\else{$e^+e^-$}\fi}
\def \to   {\mbox{$\rightarrow$}}

\def\figura#1#2#3
{\begin{figure}[hb]
  \begin{center}
  \leavevmode\protect\epsfxsize=#1\protect\epsffile{#2.eps}
  \protect\caption{#3}
  \label{#2}
  \end{center}
  \end{figure}
}
\newlength{\Largfig}
\def\({\left(}
\def\){\right)}

\def\s{\sigma}
\def\Qb{\overline{Q}}
\def\asb{{}\ifmmode \bar{\alpha}_s \else $\bar{\alpha}_s$\fi}
\def\Dh{\hat{D}}
\def\Dnp{D_{\rm NP}}

\def\OPAL{\mbox{OPAL}}
\def\ALEPH{\mbox{ALEPH}}
\def\ARGUS{\mbox{ARGUS}}

\def\ord#1{{\cal O}\(#1\)}

\relax
\def\pl#1#2#3{{\it Phys. Lett. }{\bf #1}\ (19#2)\ #3}

\def\prl#1#2#3{{\it Phys. Rev. Lett. }{\bf #1}\ (19#2)\ #3}
\def\rmp#1#2#3{{\it Rev. Mod. Phys. }{\bf#1}\ (19#2)\ #3}
\def\prep#1#2#3{{\it Phys. Rep. }{\bf #1}\ (19#2)\ #3}
\def\pr#1#2#3{{\it Phys. Rev. }{\bf #1}\ (19#2)\ #3}
\def\np#1#2#3{{\it Nucl. Phys. }{\bf #1}\ (19#2)\ #3}
\def\sjnp#1#2#3{{\it Sov. J. Nucl. Phys. }{\bf #1}\ (19#2)\ #3}
\def\app#1#2#3{{\it Acta Phys. Polon. }{\bf #1}\ (19#2)\ #3}
\def\jmp#1#2#3{{\it J. Math. Phys. }{\bf #1}\ (19#2)\ #3}
\def\nc#1#2#3{{\it Nuovo Cim. }{\bf #1}\ (19#2)\ #3}

\relax

\newcount\minutes
\newcount\scratch

\def\timestamp{%
\scratch=\time
\divide\scratch by 60
\edef\hours{\the\scratch}
\multiply\scratch by 60
\minutes=\time
\advance\minutes by -\scratch
---$\,$\hours:\null
\ifnum\minutes< 10 0\fi
\the\minutes}

\begin{document}
\begin{titlepage}
\nopagebreak
{\flushright{
        \begin{minipage}{4cm}
        Bicocca-FT-99-07 \hfill \\
        DTP/99/36 \\
        {\tt hep-ph/9903541}\hfill \\
        \end{minipage}        }

}
\vfill
\begin{center}
{\LARGE \bf \sc
 \baselineskip 0.9cm
A Phenomenological Study of\newline
Heavy-Quark Fragmentation Functions in $e^+e^-$ Annihilation

 }
\vskip 2cm
{\bf Paolo NASON}
\\
\vskip 0.1cm
{INFN, sez. di Milano, Milan, Italy} \\
\vskip .5cm
{\bf Carlo OLEARI}
\\
\vskip .1cm
{Department of Physics, University of Durham, Durham, DH1 3LE, UK}
\end{center}
\nopagebreak
\vfill
\begin{abstract}
We consider the computation of $D$ and $B$ fragmentation
functions in $\epem$ annihilation. We compare the results
of fitting present data using the next-to-leading-logarithmic
resummed approach, versus the ${\cal O}(\as^2)$ fixed-order calculation,
including also mass-suppressed effects. We also propose a method
for merging the fixed-order calculation with the resummed approach.
\end{abstract}
\vskip 1cm
March 31, 1999 \hfill
\vfill
\end{titlepage}

\section{Introduction}

A theoretical framework for the study of heavy-quark fragmentation
functions (HQFF) has been available for a long
time~\cite{MeleNason}, and several phenomenological analysis based upon this
formalism have appeared in the
literature~\cite{ColangeloNason,CacGre,RandallRius}.
In Ref.~\cite{ColangeloNason}, next-to-leading fits to the charm-momentum
spectrum at ARGUS were performed and used to predict
the bottom spectrum from $Z$ decays. From this analysis,
the non-perturbative part of the fragmentation function
predicted for the bottom quark turned out to be quite hard: in fact,
much harder than predicted in Monte Carlo models using the standard
Peterson~\cite{Peterson} parametrization. 
More precise data~\cite{ALEPH,OPAL_b,SLD} have become available since then,
giving an indication of a hard bottom fragmentation function.
In Ref.~\cite{CacGre},
a study of the charm fragmentation function was performed,
using a parametrization of the non-perturbative effects based upon the
Peterson fragmentation function, instead of the form adopted in
Ref.~\cite{ColangeloNason}. A quantitative result on the
value of the $\ep$ parameter was obtained there, definitely showing that
$\ep$  is much smaller
in next-to-leading-log (NLL) fits rather than in the leading-log (LL) ones.

The interest in further refining our understanding of the fragmentation
functions for heavy quark stems mainly from the possibility
of using them to improve
our understanding of heavy-flavour hadroproduction and photoproduction.
In Ref.~\cite{cgn98}, a formalism for the computation
of heavy-flavour production, that merges the 
high-transverse-momentum approach (i.e.\ the fragmentation-function 
approach) and the
fixed-order one, was developed. 
It was found there that mass-suppressed effects
are quite large even at moderately large transverse momenta.
This work is a first step towards a sound application of the 
fragmentation-function formalism in the moderate-transverse-momentum range.
It should, however, be complemented by similar calculations
in the context of $\epem$ annihilation, since this is the place
where the impact of non-perturbative effects is studied.
This is in fact the purpose of the present work: to use all the available
knowledge on heavy-flavour production in $\epem$ annihilation
in order to reach an assessment of the size of non-perturbative effects.

There are theoretical approaches to the fragmentation-function calculation
that rely upon heavy-quark effective theory in order to study more
systematically non-perturbative effects~\cite{Braaten}. In the present work,
however, we want to establish a connection with the most
commonly used parametrization, and thus we will use the Peterson form
throughout.  

Fixed-order (FO)
calculations of the $\ord{\as^2}$ differential cross section for
heavy-quark production in $e^+e^-$
annihilation~\cite{zbb4,Rodrigo,Bernreuther} do exist today,
and some applications in the
context of HQFF~\cite{frag97,frag98-lett} have already appeared.
In practice, the fixed-order calculation should be
more reliable than the HQFF approach for small annihilation energies.
It is interesting, therefore, to consider an approach in which
the fixed-order and the HQFF calculations are merged, in the spirit
of the work of Ref.~\cite{cgn98}, without overcounting. In the following,
we will thus review the HQFF and the fixed-order-calculation results
on fragmentation functions. We will then define a merged approach, in which
the fixed-order result is supplemented with leading and next-to-leading
logarithmically-enhanced contributions at all orders in $\as$.
We will consider both charm- and bottom-production data,
which we fit using a Peterson parametrization of the non-perturbative
contribution.

The paper is organized as follows.
In Section~\ref{sec:Theoretical} we fix our notation, and describe
the aim of our improved approach. In Section~\ref{sec:NonPerturbative}
we describe the procedure we use in order to incorporate
a parametrization of the non-perturbative effects in our formalism.
This procedure is to some extent arbitrary. It is however important
that the same procedure is used throughout our calculation.
In Section~\ref{sec:FO} we describe the fixed-order calculation
of the fragmentation function. Some subtleties, related to its
normalization, are discussed in detail.
Section~\ref{sec:NLL} is dedicated to the computation of the resummed
cross section truncated to order $\as^2$, at the NLL level (TNLL).
This calculation is almost equivalent to neglecting all mass-suppressed
terms (i.e.\ terms that vanish like powers of $m^2/E^2$) in the FO
calculation, except that terms of order $\as^2$, without any
power of $\log{E^2/m^2}$, are not included here.
In this section we spell out the definition of the improved approach,
and clarify its practical implementation.
In Section~\ref{sec:Comp} we compare the various approaches,
and in Section~\ref{sec:Fits} we describe our fits to various data sets.
Finally, in Section~\ref{sec:conclusions}, we give our conclusions.

\section{Theoretical framework}
\label{sec:Theoretical}
We consider the inclusive production of a heavy quark
$Q$ of mass $m$
\beq 
\label{eq:process}
  e^+ e^- \,\to\, Z/\gamma\;(q) \,\to\, Q\,(p) + X\;,
\eeq
where $q$ and
$p$ are the four-momenta of the intermediate boson and of the final quark.
Defining $\xe$ as the scaled energy of the final heavy quark
\beq
   \xe = \frac{2\, p\cdot q}{q^2}\;,
\eeq
and introducing the centre-of-mass energy $E=\sqrt{q^2}$, we have the
physical constraint
\beq
 \sqrt{\rho} \le \xe \le 1\;,
\eeq
where
\beq
\rho= \frac{4\,m^2}{E^2}\;.
\eeq
At times, we will instead use the normalized momentum fraction, defined as
\beq
\label{eq:momentumfrac}
 \xp = \frac{\sqrt{\xe^2 - \rho}}{\sqrt{1-\rho}}\;,\quad\quad
   0 \le \xp \le 1 \:.
\eeq
The inclusive cross section for the production of a heavy quark 
can be written as a perturbative expansion in $\as$
\beq
\label{eq:sigma}
   \frac{d\sigma}{d\xp}(\xp,E,m)
= \sum_{n=0}^{\infty} a^{(n)}(\xp,E,m,\mu)\, \asb^n(\mu)\;,
\eeq
where $\mu$ is the renormalization scale, and
\beq
 \asb(\mu) =  \frac{\as(\mu)}{2\pi}\;.
\eeq

If $\mu \approx E \approx m$, the truncation of Eq.~(\ref{eq:sigma}) 
at some fixed
order in the coupling constant can be used to approximate the cross
section.  An $\ord{\as^2}$ fixed-order calculation for the
process~(\ref{eq:process}) is
available~\cite{zbb4,Rodrigo,Bernreuther}, so that we can compute the
coefficients of Eq.~(\ref{eq:sigma}) at the $\ord{\as^2}$ level.
We thus define the fixed-order  result as
\beq
\label{eq:sigma_fix}
 \left.  \frac{d\sigma}{d\xp}(\xp,E,m)\fix
= a^{(0)}(\xp,E,m) + a^{(1)}(\xp,E,m)\, \asb(E) + a^{(2)}(\xp,E,m)\,
 \asb^2(E) \;,
\eeq
where we have taken $\mu=E$ for ease of notation.

If $E\gg m$, large logarithms of the form $\log\(E^2/m^2\) $ appear
in the differential cross section~(\ref{eq:sigma}) to
all orders in the perturbative expansion. In this limit, if we disregard all
power-suppressed terms of the form $m^2/E^2$,
the inclusive cross section can be organized in the expansion
\beqn
\label{eq:resummed_sigma}
  \left. \frac{d\sigma}{dx}(x,E,m) \res \!&=&\!
\sum_{n=0}^{\infty} \beta^{(n)}(x) \,\( \asb(E) \log\frac{E^2}{m^2}
\)^n \!+ \asb(E)\sum_{n=0}^{\infty} \gamma^{(n)}(x) \,\( \asb(E)
\log\frac{E^2}{m^2}\)^n
\nn \\
 && {} + \asb^2 (E) \sum_{n=0}^{\infty}  \delta^{(n)}(x) \,
 \( \asb(E)\log\frac{E^2}{m^2}\)^n + 
\ldots + \ord{\frac{m^2}{E^2}}\;,
\eeqn
where $x$ stands now for either $\xe$ or $\xp$, since the two variables
differ by power-suppressed effects.

We define the leading-logarithmic (LL) approximation as
\beq
\label{eq:LL}
\left. \frac{d\sigma}{dx}(x,E,m) \LL  =
\sum_{n=0}^{\infty} \beta^{(n)}(x) \,\( \asb(E) \log\frac{E^2}{m^2}
\)^n \;,
\eeq
and the next-to-leading-logarithmic (NLL) one as
\begin{equation}
\label{eq:resummed}
  \left. \frac{d\sigma}{dx}(x,E,m) \NLL \! =
\sum_{n=0}^{\infty} \beta^{(n)}(x) \,\( \asb(E) \log\frac{E^2}{m^2}
\)^n \!\!+ \asb(E)\sum_{n=0}^{\infty} \gamma^{(n)}(x) \,\( \asb(E)
\log\frac{E^2}{m^2}\)^n\,.
\end{equation}
The $\delta^{(n)}$ coefficients define the NNLL terms, that are, as of
now, not known.

The expansion of Eq.~(\ref{eq:resummed}) up to order $\as^2$ is given by
\beqn
\left. \frac{d\sigma}{dx}(x,E,m) \NLL \!
&=&\!\beta^{(0)}(x) + \asb(E) \( \gamma^{(0)}(x) + \beta^{(1)}(x)
\log\frac{E^2}{m^2}\) \nn\\
&&\!\mbox{} + \asb^2(E) \( \gamma^{(1)}(x)\log\frac{E^2}{m^2} + 
\beta^{(2)}(x) \log^2\frac{E^2}{m^2} \) + \ord{\as^3}\;,\phantom{aaaa}
\eeqn
and does not coincide with the massless limit of the FO calculation.
A term of order $\as^2$, not accompanied by logarithmic factors,
may in fact survive in the massless limit of the FO result.  In
the HQFF approach, this is a NNLL effect, and therefore it is not included
at the NLL level. We will refer to this term, in the following, as
the NNLL $\as^2$ term.

It is now clear how to obtain an improved formula, which contains all the
information present in the FO approach, as well as in the HQFF approach.
Using Eqs.~(\ref{eq:sigma_fix}) and~(\ref{eq:resummed}), we write the
improved cross section as 
\beqn
\label{eq:improved}
   \left. \frac{d\sigma}{d\xp}(\xp,E,m) \imp &= &
\sum_{i=0}^2
a^{(i)}(\xp,E,m)\,\asb^i(E)
 +\sum_{n=3}^{\infty} \beta^{(n)}(x) \,\( \asb(E) \log\frac{E^2}{m^2}
\)^n 
\nn\\
&&{}+\asb(E)\sum_{n=2}^{\infty} \gamma^{(n)}(x) \,\( \asb(E)
\log\frac{E^2}{m^2}\)^n \;,
\eeqn
where the LL and NLL sums now start from $n=3$ and $n=2$
respectively, in order to avoid double counting.
Formula (\ref{eq:improved}) includes exactly all
terms up to the order $\as^2$ (including mass effects), and all terms of the
form $\Big( \asb(E) \log\(E^2/m^2\)\Big)^n $ and $\asb(E) \, \Big( \asb(E)
\log\(E^2/m^2\) \Big)^n $, so that it is also
correct at NLL level for $E \gg m$. It can be viewed as an
interpolating formula. For moderate energies, it is accurate to the order
$\as^2$, while for very large energies it is accurate at the NLL level.

\section{Non-perturbative effects}
\label{sec:NonPerturbative}

In the HQFF formalism, the inclusive heavy-flavour cross section
is given by the formula
\beqn
\label{eq:factorization}
   \frac{d\s}{dx}(x,E,m) &=& 
 \sum_i  \frac{d\hat\s_i}{dx}\(x,E,\mu_F\) \,
 \otimes \Dh_i\(x,\mu_F,m\) \nn\\
&=& \sum_i \int_x^1 \frac{dz}{z}\, 
 \frac{d\hat\s_i}{dz}\(z,E,\mu_F\) \,
  \Dh_i\(\frac{x}{z},\mu_F,m\) \;,
\eeqn
where $d\hat\s_i(x,E,\mu_F)/dx$ are the \MSB-subtracted partonic cross
sections for producing the parton $i$, and $\Dh_i(x,\mu_F,m)$ are the
\MSB\ fragmentation functions for parton $i$ to evolve
into the heavy quark $Q$. The factorization scale
$\mu_F$  must be taken of order $E$ in
order to avoid the appearance of large logarithms of $E/\mu_F$ in the
partonic cross section.  The explicit expressions for the partonic
cross sections and for the fragmentation functions at NLO can be found
in Refs.~\cite{MeleNason,NasonWebber}.

The weak point of formula (\ref{eq:factorization}) comes from the
initial condition for the evolution of the fragmentation function,
which is computed as a power expansion in terms of $\as(m)$.
In particular, irreducible, non-perturbative uncertainties of
order $\LambdaQCD/m$ are present.
We assume that all these
effects are described by a non-perturbative fragmentation function
$\Dnp^H$, that takes into account all low-energy effects, including
the process of the heavy quark turning into a
heavy-flavoured hadron. The full resummed cross section, including
non-perturbative corrections, is then written as 
\beqn
\label{eq:hadfactor}
   \frac{d\s^H}{dx}(x,E,m) &=& 
\sum_i  \frac{d\hat\s_i}{dx}\(x,E,\mu_F\) \,
 \otimes \Dh_i\(x,\mu_F,m\) \otimes \Dnp^H (x) \;.
\eeqn
We will parametrize the non-perturbative part of the
fragmentation function with the Peterson form
\beq
\label{eq:peterson}
\Dnp^H (x) = P(x,\ep) \equiv 
 N \,\frac{x\,(1-x)^2}{\lq (1-x)^2+x\,\ep\rq^2} \;,
\eeq
where the normalization factor $N$ determines
the fraction of the hadron of type $H$ in the final state.
Summing over all hadron types, we have the condition
\beq
\sum_H  \int_0^1  dx \, \Dnp^H (x) = 1\;.
\eeq

\section{The $\as^2$ fixed-order approach}
\label{sec:FO}

We define our fixed-order cross section, supplemented with non-perturbative
fragmentation effects, by the
convolution of the perturbative cross section~(\ref{eq:sigma_fix})
with the non-perturbative fragmentation function~(\ref{eq:peterson})
\beqn
\label{eq:sig_fix_had}
 \left. \frac{d\s^H}{d\xp}(\xp,E,m)\fix &=& \int_0^1 dy\,d\zp
 \left. \frac{d\s}{d\zp}(\zp,E,m)\fix  P(y,\ep) \,\delta(\xp - y\zp) \nn\\
 & = & \int_0^1 dy\,d\zp \sum_{i=0}^2 a^{(i)}(\zp,E,m)\,\asb^i(E)\;
 P(y,\ep)\,  \delta(\xp - y\zp)\;.
\eeqn
It is assumed, in this formula, that non-perturbative effects
degrade the momentum, rather than the energy of the heavy quark.
This is to avoid unphysical results when the momentum is smaller than
the mass. It should be made clear, however, that for small momenta
this formula should be seen merely as a model for non-perturbative effects.

The coefficients of the perturbative expansion $a^{(i)}(\zp,E,m)$ are in
general distributions in $\zp$ with singularities at $\zp=1$. In order to deal
with these singularities, we transform Eq.~(\ref{eq:sig_fix_had}) as follows
\beqn
\label{eq:sig_fix_final}
 \left. \frac{d\s^H}{d\xp}\fix 
 &=&  \int_0^1 dy\,d\zp \sum_{i=0}^2 a^{(i)}(\zp,E,m)\,\asb^i(E)\;
  P(y,\ep)\, \lg \delta(\xp - y\zp) - \delta(\xp-y) \rg \nn\\
&+&  \int_0^1 dy\,d\zp \sum_{i=0}^2 a^{(i)}(\zp,E,m)\,\asb^i(E)\;
  P(y,\ep)\, \delta(\xp - y)  \nn\\
 &=&  P(\xp,\ep)  \sum_{i=0}^2 \bar{a}^{(i)}(E,m)\,\asb^i(E) \nn\\
&+& \int_0^1 dy\,d\zp \sum_{i=0}^2 a^{(i)}(\zp,E,m)\,\asb^i(E)\;
 P(y,\ep)\, \lg\delta(\xp - y\zp) - \delta(\xp-y) \rg , \phantom{aaaa}
\eeqn
where
\beq
 \bar{a}^{(i)}(E,m) \equiv \int_0^1 d\zp \, a^{(i)}(\zp,E,m)\;.
\eeq
The $a^{(0)}$ term in the last integral of Eq.~(\ref{eq:sig_fix_final}),
in fact, does not contribute, since it is proportional to
 $\delta(1-\zp)$. We thus find 
\beqn
\label{eq:sig_fix_had_mod}
 \left. \frac{d\s^H}{d\xp}\fix 
 &=& \!\! P(\xp,\ep) \,\s_{\rm inc} \nn\\
 &+& \!\!\int_0^1 dy\,d\zp  \sum_{i=1}^2 a^{(i)}(\zp,E,m)\,\asb^i(E)\;
 P(y,\ep)\, \lg\delta(\xp - y\zp) - \delta(\xp-y) \rg \,,\phantom{aaa}
\eeqn
where
\beq
\label{eq:sig_tot_as2}
\s_{\rm inc} =  \bar{a}^{(0)}(E,m) +\bar{a}^{(1)}(E,m)\,\asb(E) +
 \bar{a}^{(2)}(E,m)\, \asb^2(E)\;
\eeq
is the inclusive heavy-quark cross section.

It would be natural to normalize the cross section in terms of
the total cross section for heavy-flavoured events. However, for
practical purposes, the normalization is immaterial, because of
the uncertainty in the specific heavy-flavoured hadron fraction, and it
must be fitted to the data. It is important,
however, that the normalization factor has a finite massless limit,
since the same normalization has to be applied to the resummed cross section.
In the present case,
neither the total heavy-flavour cross section nor the inclusive
cross section are finite in the massless limit.
This problem arises because of mass singularities in the coefficient
$\bar{a}^{(2)}(E,m)$.
We then define
\beq
\bar{a}^{(2)}(E,m)= \bar{a}_{1Q}^{(2)}(E,m)+2\,
\bar{a}_{2Q}^{(2)}(E,m)+\bar{a}_{R}^{(2)}(E,m)\;,
\eeq
where:
\begin{itemize}
\item[-]
the $1Q$-term consists of all graphs in which the primary interaction vertex
is always attached to the heavy flavour, and furthermore there is
only one heavy-flavour pair in the final state.
\item[-]
The $2Q$-term  consists of all graphs in which the primary interaction vertex
is always attached to the heavy flavour, and there is a secondary
heavy-quark pair in the final state coming from the gluon-splitting mechanism.
There is a factor of 2 in front of this contribution, since, in the
inclusive cross section, both the primary heavy quark and the secondary one
can be detected.
\item[-]
The $R$-term (where $R$ stands for ``rest'') contains all other contributions.
These include terms in which a heavy-flavour pair is produced via gluon
splitting in a process initiated by light quarks, plus interference terms
in which a heavy quark (antiquark), produced via gluon splitting in an
amplitude, interferes with a heavy quark (antiquark), produced directly
at the $Z/\gamma$ vertex.
\end{itemize}
We define a ``normalization'' total cross section as
\beq
 \s_{\rm n} =  \bar{a}^{(0)}(E,m) +\bar{a}^{(1)}(E,m)\,\asb(E) +
 \left[\bar{a}_{1Q}^{(2)}(E,m)+\bar{a}_{2Q}^{(2)}(E,m)\right]\, \asb^2(E)\;.
\eeq
This cross section is finite in the massless limit. In fact, the mass
singularities, present in the $\bar{a}_{2Q}^{(2)}(E,m)$ term, cancel
against the virtual graphs in $\bar{a}_{1Q}^{(2)}(E,m)$
having a gluon self-energy correction consisting of a heavy-flavour loop.

The analytic expressions of  $a^{(0)}(\zp,E,m)$ and $a^{(1)}(\zp,E,m)$,
and of $\bar{a}^{(0)}(E,m)$ and $\bar{a}^{(1)}(E,m)$,
are known (see Refs.~\cite{NasonWebber,Reinders}).
The $a^{(2)}(\zp,E,m)$ term has been computed by the authors and is
available as a numerical FORTRAN program.  

The advantage of writing
Eq.~(\ref{eq:sig_fix_had}) in the form of
Eq.~(\ref{eq:sig_fix_had_mod}) is that
the singularities in the two-jet limit are regularized,
and the two-body virtual terms (i.e.\ $Z/\gamma \,\to\, Q +
\Qb$), that are proportional to $\delta(1-\zp)$, give zero contribution.
Thus, the two-loop virtual
corrections, which were not computed in Ref.~\cite{zbb4}, do not contribute
there.
On the other hand, they contribute to $\bar{a}^{(2)}_{1Q}(E,m)$.
They do disappear from our formula, however,
if we normalize it to the $\s_{\rm n}$ cross
section, consistently dropping higher order terms. We obtain
\newcommand\nr{n_R}
\beqn
\frac{1}{\s_{\rm n}} \left. \frac{d\s^H}{d\xp}(\xp,E,m)\fix 
  &=&   P(\xp,\ep)\,\nr
+ \frac{1}{\bar{a}^{(0)}(E,m)}
\int_0^1 dy\,d\zp \,\Bigg\{  a^{(1)}(\zp,E,m)\,\asb
\nn\\
&&{}+  \lq a^{(2)}(\zp,E,m)  -\frac{\bar{a}^{(1)}(E,m)}{\bar{a}^{(0)}(E,m)}
\, a^{(1)}(\zp,E,m) \rq
 \asb^2 \Bigg\} 
\nn\\
&&{}\times
P(y,\ep)\, \lg\delta(\xp - y\zp) - \delta(\xp-y)
 \rg \;,
\eeqn
where
\beq
\nr=1+\frac{\bar{a}_{2Q}^{(2)}(E,m)+\bar{a}_{R}^{(2)}(E,m)}{\bar{a}^{(0)}(E,m)}
  \,\asb^2(E)\;.
\eeq
Although the $\bar{a}_{R}^{(2)}(E,m)$ term contains all sort of interference
contributions, it is dominated by terms in which a primary light-quark
pair generates a gluon that decays into a heavy-quark pair.
This contribution is singular in the massless limit.
An analogous singularity is present in
the $\bar{a}_{2Q}^{(2)}(E,m)$ term. Thus, the dominant corrections
to $\nr$ arise from secondary heavy-quark production via gluon splitting.

Performing the integration in $y$ we can write
\beqn
\label{eq:sig_fix_norm}
\frac{1}{\s_{\rm n}}\left. \frac{d\s^H}{d\xp}(\xp,E,m)\fix 
  &=&   P(\xp,\ep)\,\nr \nn\\
&& {}\hspace{-1.3cm}+ \sum_{i=1}^{2} \, \asb^i (E) \,
\Biggl\{ \int_{\xp}^1 d\zp\,  b^{(i)}(\zp,E,m) 
\lq \frac{P\(\xp/\zp,\ep\)}{\zp}  - P(\xp,\ep) \rq \nn\\
&&{}\hspace{-1.3cm}- P(\xp,\ep) \int_{0}^{\xp} d\zp\,
b^{(i)}(\zp,E,m) \Biggr\} \;, 
\eeqn
where we have defined
\beqn
 b^{(1)}(\zp,E,m) &=& \frac{ a^{(1)}(\zp,E,m)}{\bar{a}^{(0)}(E,m)}\nn\\
 b^{(2)}(\zp,E,m) &=& \frac{1}{\bar{a}^{(0)}(E,m)} 
\lq a^{(2)}(\zp,E,m)  -\frac{\bar{a}^{(1)}(E,m)}{\bar{a}^{(0)}(E,m)}
\, a^{(1)}(\zp,E,m) \rq \;. \phantom{aaa}
\eeqn

For future use, we introduce here the hadronic differential cross section
expressed in term of the energy fraction. From
Eq.~(\ref{eq:momentumfrac}), we can write 
\beq 
\label{eq:sigmahad_xe}
\frac{1}{\s_{\rm n}}\, \frac{d\s^H}{d\xe} =
\frac{1}{\s_{\rm n}}\, \frac{d\s^H}{d\xp} \,\frac{d\xp}{d\xe} =
\frac{1}{\s_{\rm n}} \,\frac{d\s^H}{d\xp}\, \frac{\xe}{\xp}\,
\frac{1}{1-\rho} \;.
\eeq

\section{NLL higher-order effects}
\label{sec:NLL}

According to Eq.~(\ref{eq:improved}), we must add to the FO
result all terms of order $\as^3$ and higher of the full NLL
resummed result. The resummed result is obtained numerically
by solving the Altarelli-Parisi evolution equations for
the fragmentation functions, with the given ${\cal O}(\as)$ initial
conditions, convoluted with the appropriate short-distance cross sections.
We must, however, subtract, from this numerical result, the terms up to
second order of its expansion in powers of $\as$.
These terms can be obtained with the same procedure used in
Ref.~\cite{frag97}.   
We recall
here the principal steps that give us the $\ord{\as^2}$ terms of the
resummed cross section.

We introduce the following notation for the Mellin transform of a
generic function $f(x)$:
\beq
   f(N) \equiv \int_0^1 dx \, x^{N-1} f(x)\;.
\eeq
We adopt the convention that, when $N$ appears, instead of $x$, as the
argument of a function, we are actually referring to the Mellin transform
of the function. This notation is somewhat improper, but it should
not generate confusion in the following, since we will work
only with Mellin transforms. 

The Mellin transform of the factorization
theorem~(\ref{eq:factorization}) is given by
\beq\label{eq:sigmaN}
\sigma(N,E,m) =\sum_i \hat\sigma_i(N,E,\mu) \; \Dh_i(N,\mu,m)\;,
\eeq
where
\beq
\sigma(N,E,m) = \int_0^1 dx \, x^{N-1} \frac{d\s}{dx}(x,E,m)\;,
\eeq
and a similar one for $\hat{\s}_i(N,E,\mu)$. The Mellin transform
of the Altarelli-Parisi evolution equations is
\beq
\label{eq:APN}
 \frac{d\Dh_i(N,\mu,m)}{d\log\mu^2} = \sum_j 
 \asb(\mu) \left[ P_{ij}^{(0)}(N) + P_{ij}^{(1)}(N)\,
     \asb(\mu) +   {\cal O}\(\as^2\)\right] \Dh_j(N,\mu,m)\;.
\eeq
We need  an expression for $\sigma(N,E,m)$ valid at the second
order in $\as$.
Thus, we solve Eq.~(\ref{eq:APN}), with initial condition
at $\mu=\mu_0$, with an $\ord{\as^2}$ accuracy.
This is easily done by rewriting
Eq.~(\ref{eq:APN}) as an integral equation
\beqn
 \Dh_i(N,\mu,m)\!\!\! &=& \!\!\!\Dh_i(N,\mu_0,m) \nn \\ 
\label{eq:integAPN0} 
\!\!\! &+& \!\!\! \sum_j \int_{\mu_0}^\mu d\log{\mu^\prime}^2\;
 \asb(\mu^\prime) \left[ P_{ij}^{(0)}(N) + P_{ij}^{(1)}(N)\,
     \asb(\mu^\prime) \right]  \Dh_j(N,\mu^\prime,m)
\;. \phantom{aaaaa}
\eeqn
The terms proportional to $\as^2$ can be evaluated at any scale
($\mu$ or $\mu_0$),
the difference being of order $\as^3$. Factors involving a single
power of $\as$ can instead be expressed in terms of $\as(\mu_0)$
using the renormalization group equation
\beqn
\label{eq:RGE}
\asb(\mu^\prime) &=& \asb(\mu_0)-2\,\pi\,b_0\,\asb^2(\mu_0)\,
\log\frac{{\mu^\prime}^2}{\mu_0^2}+{\cal O}\(\as^3\(\mu_0\)\) \nn\\
 b_0 &=& \frac{11\,C_A - 4\, n_f\,T_F}{12\,\pi}\;,
\eeqn
with $n_f$ the number of flavours, including the heavy one.
Equation~(\ref{eq:integAPN0}) then becomes
\beqn
 \Dh_i(N,\mu,m) &=& \Dh_i(N,\mu_0,m)
+ \sum_j \int_{\mu_0}^\mu d\log{\mu^\prime}^2\;
 \asb(\mu_0) P_{ij}^{(0)}(N) \,
 \Dh_j(N,\mu^\prime,m)
 \nonumber \\  
&&{}+ \sum_j  \asb^2(\mu_0)\, P_{ij}^{(1)}(N) \,
 \Dh_j(N,\mu_0,m)\,\log\frac{\mu^2}{\mu_0^2}  \nonumber \\ 
\label{eq:integAPN1} 
&& {} -  2\,\pi\,b_0\,\sum_j
 \asb^2(\mu_0)\, P_{ij}^{(0)}(N)\,
 \Dh_j(N,\mu_0,m)\,\frac{1}{2}\log^2\frac{\mu^2}{\mu_0^2}  \;.
\eeqn
Now we  need to express $\Dh_j(N,\mu^\prime,m)$ on the right-hand side
of the above equation as a function of the initial condition,
with an accuracy of order $\as$. This is simply done by iterating
the above equation once, keeping only the first two terms on the 
right-hand side. Our final result is then
\beqn
 \Dh_i(N,\mu,m) &=& \Dh_i(N,\mu_0,m)
+ \sum_j
 \asb(\mu_0) P_{ij}^{(0)}(N) \, 
 \Dh_j(N,\mu_0,m) \, \log\frac{\mu^2}{\mu_0^2}
 \nonumber \\ 
&&{} +  \sum_{kj}
 \asb^2(\mu_0) P_{ik}^{(0)}(N) \, P_{kj}^{(0)}(N)\,
 \Dh_j(N,\mu_0,m) \, \frac{1}{2}\log^2\frac{\mu^2}{\mu_0^2}
 \nonumber \\  
&&{} +  \sum_j \asb^2(\mu_0) P_{ij}^{(1)}(N) \,
 \Dh_j(N,\mu_0,m) \, \log\frac{\mu^2}{\mu_0^2}
 \nonumber \\ \label{eq:integAPN2} 
&&{} -  \pi\,b_0\,\sum_j
 \asb^2(\mu_0) P_{ij}^{(0)}(N) \, 
 \Dh_j(N,\mu_0,m) \, \log^2\frac{\mu^2}{\mu_0^2} \;.
\eeqn
Using the initial conditions (see Ref.~\cite{MeleNason})
\beqn
\Dh_Q(N,\mu_0,m) &=& 1 +\asb(\mu_0)\,d^{(1)}_Q(N,\mu_0,m)
   +{\cal O}\(\as^2\(\mu_0\)\)\nn\\ 
\label{eq:ini}
\Dh_g(N,\mu_0,m) &=& \asb(\mu_0)\,d^{(1)}_g(N,\mu_0,m)
   +{\cal O}\(\as^2\(\mu_0\)\)\;,
\eeqn
(all the other components being of order $\as^2$),
and using
Eq.~(\ref{eq:RGE}) to express $\as\(\mu_0\)$ in terms of $\as\(\mu\)$,
Eq.~(\ref{eq:integAPN2}) becomes, with the required accuracy,
\beqn
 \Dh_i(N,\mu,m) &=& \delta_{iQ}+\asb(\mu)\,d^{(1)}_i(N,\mu_0,m)
+ 2\, \pi\,b_0\,\asb^2(\mu)\,d^{(1)}_i(N,\mu_0,m)\,\log\frac{\mu^2}{\mu_0^2}
 \nonumber \\ 
&&{} + \asb(\mu) P_{iQ}^{(0)}(N) \, \log\frac{\mu^2}{\mu_0^2}
+ \sum_j \asb^2(\mu) P_{ij}^{(0)}(N) \, 
 d^{(1)}_j(N,\mu_0,m) \, \log\frac{\mu^2}{\mu_0^2}
 \nonumber \\ 
&& {} +  \sum_{k} \asb^2(\mu) P_{ik}^{(0)}(N) \, P_{kQ}^{(0)}(N)
 \, \frac{1}{2}\log^2\frac{\mu^2}{\mu_0^2}
+ \asb^2(\mu) P_{iQ}^{(1)}(N)\, \log\frac{\mu^2}{\mu_0^2}
 \nonumber \\ 
\label{eq:integAPN} 
&&{} +  \pi\,b_0\, \asb^2(\mu) P_{iQ}^{(0)}(N)
 \, \log^2\frac{\mu^2}{\mu_0^2} \;.
\eeqn
Using the notation of Ref.~\cite{NasonWebber}, we have, for the 
partonic cross sections, 
\beqn
\sh_Q &=& \sh_{\Qb} = \s_{0,Q}\lq 1+\asb C_F c_Q + \ord{\as^2}\rq \nn\\
\sh_q &=& \sh_{\qb} = \s_{0,q}\lq1+\ord{\as}\rq \nn\\
\label{eq:sigmahat}
\sh_g &=& \s_{0} \lq \asb C_F c_g +\ord{\as^2}\rq\;,
\eeqn
where $Q$ is the heavy quark and $q$ is a light one, and
\beqn
\s_{0,Q} &=& \s_{0,Q}^{(v)}+\s_{0,Q}^{(a)}\nn\\
\s_{0,q} &=& \s_{0,q}^{(v)}+\s_{0,q}^{(a)}\nn\\
\s_{0} &=& \sum_{f=1}^{n_f} \s_{0,f} = \s_{0,Q} + \sum_{q} \s_{0,q} \nn\\
c_Q &=& c_{T,Q} + c_{L,Q} \nn\\
c_g &=& c_{T,g} + c_{L,g}\;,
\eeqn
where $v$, $a$, $T$ and $L$ refer to the vector, axial, transverse and 
longitudinal contributions, respectively.

In addition, we define the Mellin moments
\beq
\ah_{Q/g}^{(1)}(N,E,\mu)= C_F c_{Q/g}(N,E,\mu)\;,
\eeq
and introduce the total ``normalization'' $\ord{\as}$
cross section for the production of a heavy quark
\beq
\label{eq:sig_norm}
 \s_{\rm n} =
 \s_{0,Q}\lq 1+ \frac{3}{2} C_F \asb  + \ord{\as^2}\rq 
 \;.
\eeq
Observe that now, at NLL accuracy, we can neglect the $\ord{\as^2}$
terms, since they do not contain any logarithmic enhancement.
This would not be the case for the total heavy-quark cross section,
that in fact is divergent at order $\ord{\as^2}$, in the massless
limit approximation.

From Eq.~(\ref{eq:sigmaN}), we can write the NLL cross section as
\beq
 \s(N) = 
\sh_Q \,\Dh_Q + \sh_{\Qb}\,\Dh_{\Qb} + \sum_q \sh_q \, \Dh_q  + \sum_{\qb}
 \sh_{\qb}\,\Dh_{\qb}  + \Dh_g \,\sh_g\;,
\eeq
where we have dropped the energy and the mass dependence, for ease of
notation.

We can now  obtain the truncated $\ord{\as^2}$ NLL (TNLL from now on)
normalized cross section as
\beqn
\left.\frac{1}{\s_{\rm n}}\s(N)\right|_{\rm TNLL}\!\!\!\!\!\!\!\!
&=& \!\!\!\Biggl\{ 
1+\asb(\mu) \,d^{(1)}_Q(N)  + \asb(\mu) P_{QQ}^{(0)}(N) \, \L + \asb(\mu)
\left(\ah_Q^{(1)}(N)-\frac{3}{2}C_F \right)  \nn\\
&&{}\!\!\! + \asb^2(\mu) P_{QQ}^{(0)}(N) \left(\ah_Q^{(1)}(N)
-\frac{3}{2}C_F \right) \L 
+ 2\, \pi\,b_0\,\asb^2(\mu)\,d^{(1)}_Q(N) \, \L \nn\\ 
&&{}\!\!\! 
+ \sum_j \asb^2(\mu) P_{Qj}^{(0)}(N) \, 
 d^{(1)}_j(N) \, \L  \nn \\ 
&& {} +  \sum_{k} \asb^2(\mu) P_{Qk}^{(0)}(N) \, P_{kQ}^{(0)}(N)
 \, \frac{1}{2} \log^2\frac{\mu^2}{\mu_0^2}
+ \asb^2(\mu) P_{QQ}^{(1)}(N)\, \L
 \nonumber \nn\\ 
&&{}\!\!\! +  \pi\,b_0\, \asb^2(\mu) P_{QQ}^{(0)}(N)
 \, \log^2\frac{\mu^2}{\mu_0^2} \nn\\
&&{}\!\!\!+  \asb^2(\mu) P_{\Qb g}^{(0)}(N) \, 
 d^{(1)}_g(N) \, \log\frac{\mu^2}{\mu_0^2}
+  \asb^2(\mu) P_{\Qb g}^{(0)}(N) \, P_{gQ}^{(0)}(N)
 \, \frac{1}{2}\log^2\frac{\mu^2}{\mu_0^2}\nn\\
&&{}\!\!\!+ \asb^2(\mu) P_{\Qb Q}^{(1)}(N)\, \log\frac{\mu^2}{\mu_0^2} \Biggr\} 
\nn \\  
&&{}\!\!\! +\frac{1}{\sh_{0,Q}}\sum_q \sh_{0,q} \asb^2(\mu) \Biggl\{
 P_{qg}^{(0)}(N) \, 
 d^{(1)}_g(N) \, \log\frac{\mu^2}{\mu_0^2}
 \nonumber \\ 
&&{}\!\!\! +   P_{qg}^{(0)}(N) \, P_{gQ}^{(0)}(N)
 \, \frac{1}{2}\log^2\frac{\mu^2}{\mu_0^2}
+ P_{qQ}^{(1)}(N)\, \log\frac{\mu^2}{\mu_0^2}
 \nonumber \\ 
&&{}\!\!\! + P_{\qb g}^{(0)}(N) \, 
 d^{(1)}_g(N) \, \log\frac{\mu^2}{\mu_0^2}
 \nonumber \\ 
&&{}\!\!\! +    P_{\qb g}^{(0)}(N) \, P_{gQ}^{(0)}(N)
 \, \frac{1}{2}\log^2\frac{\mu^2}{\mu_0^2}
+ P_{\qb Q}^{(1)}(N)\, \log\frac{\mu^2}{\mu_0^2} \Biggr\} \nn\\ 
&&{}\!\!\! + \frac{1}{\sh_{0,Q}} \sh_{0} \ah_g^{(1)}(N)
  \asb^2(\mu) P_{gQ}^{(0)}(N) \,
\log\frac{\mu^2}{\mu_0^2} \;,
\eeqn
where we have taken $\mu=E$ and $\mu_0=m$, and we have introduced the
notation
\beq
\hat{a}^{(1)}_{Q/g}(N)=\hat{a}^{(1)}_{Q/g}(N,E,\mu)\vert_{\mu=E}\;,
\quad\quad\quad 
d^{(1)}_{Q/g}(N)=d^{(1)}_{Q/g}(N,\mu_0,m)\vert_{\mu_0=m}\;.
\eeq
The lowest order splitting functions are given by
\beqn
P^{(0)}_{QQ}(N) &=& 
  C_F\left[\frac{3}{2}+\frac{1}{N(N+1)}-2\,S_1(N)\right]\,,
\nonumber \\
P^{(0)}_{Qg}(N) &=& P^{(0)}_{\Qb g}(N) = P^{(0)}_{qg}(N) = P^{(0)}_{\qb g}(N)
=  C_F\left[\frac{2+N+N^2}{N(N^2-1)}\right]\,,
\nonumber \\
P^{(0)}_{gQ}(N) &=& T_F\left[ \frac{2+N+N^2}{N(N+1)(N+2)}\right]\,,
\eeqn
where
\beq
S_1(N)=\psi_0(N+1)-\psi_0(1)\,,
\eeq
and 
\beq
\psi_0(z) = \frac{d}{dz}\log{\Gamma(z)}\;.
\eeq
In addition we have
\beq
P_{q Q}^{(1)}(N) = P_{\qb Q}^{(1)}(N) = P_{q^\prime q}(N)\;,
\eeq
and the splitting functions $P_{QQ}^{(1)}(N)$, $P_{\Qb Q}^{(1)}(N)$ and
$P_{q^\prime q}(N)$, together with the initial condition for the fragmentation
functions $d^{(1)}_{Q}(N)$ and $d^{(1)}_{g}(N)$, can be found in
Refs.~\cite{MeleNason,frag97}.

The differential cross section, as a function of $x$, can now
be obtained by an inverse Mellin transform.
We define
the hadronic cross section, including non-perturbative effects
\beq \label{eq:TNLL}
\left. \frac{1}{\sigma_{\rm n}}\frac{d\sigma^H}{dx}(x,E,m) \TNLL =
  \frac{1}{2\pi i} \int_{c-i\infty}^{c+i\infty} dN\, 
\left.\frac{1}{\sigma_{\rm n}} \sigma(N)\TNLL P(N,\ep) 
\, x^{-N} \;,
\eeq
where $P(N,\ep)$ is the Mellin transform of the Peterson function.
We can now isolate the higher-order effects (HOE)
needed in the expression of the improved
cross section of Eq.~(\ref{eq:improved}). They can be
computed as a difference between the NLL and the TNLL cross section
\beqn
{\rm (HOE)} &\equiv& \Bigg\{
\sum_{n=3}^{\infty} \beta^{(n)}(x) \,\( \asb(E) \log\frac{E^2}{m^2} \)^n 
\nn\\
&&{}+\asb(E)\sum_{n=2}^{\infty} \gamma^{(n)}(x) \,\( \asb(E)
\log\frac{E^2}{m^2}\)^n \Bigg\} \otimes P(x,\ep)  \nn\\
&=& \left. \frac{d\sigma^H}{dx}(x,E,m) \NLL -
    \left. \frac{d\sigma^H}{dx}(x,E,m) \TNLL \;,
\eeqn
where
\beq
\label{eq:sigNLLH}
\left. \frac{d\sigma^H}{dx}(x,E,m) \NLL = 
  \left. \frac{d\sigma}{dx}(x,E,m) \NLL \otimes P(x,\ep)\;.
\eeq

We can then summarize our improved cross section as
\beq
\label{eq:improvedH}
\frac{1}{\s_{\rm n}} \left. \frac{d\s^H}{d\xp}\imp\  = 
\frac{1}{\s_{\rm n}} \left. \frac{d\s^H}{d\xp}\fix\  +
 \frac{1}{\s_{\rm n}} \left. \frac{d\s^H}{dx}\NLL\  -
\frac{1}{\s_{\rm n}} \left. \frac{d\s^H}{dx}\TNLL \;,
\eeq
or a similar one with $\xp$ replaced by $\xe$.

\section{Comparison of the various approaches} 
\label{sec:Comp}

In this section we perform a comparison of the different approaches
presented so far.
We confine ourselves to $D$ and $B$ meson production at $E=10.6$~GeV
and $E=91.2$~GeV, which are relevant to the data sets that
we will fit in the forthcoming section, where we analyze the experimental
results obtained by \ARGUS\ and \OPAL\ for $D$ mesons, and by \ALEPH\
for $B$ mesons.
We fix the charm and bottom mass to 1.5 and 5~GeV, respectively.
For $D$ meson production we take $\ep_D=0.035$, while
for $B$ mesons we use $\ep_B=(m_c^2/m_b^2) \, \ep_D\approx 0.0035$.
The renormalization and factorization scales have been settled
equal to the total energy $E$.
Furthermore, we have fixed $\Lambda_{\rm QCD}^{(5)} = 200$~MeV,
so that $\as^{(5)}(M_Z)=0.116$.

In the fixed-order calculation, we have taken three light flavours, for the
energy of $10.6$~GeV, and four light flavours for $E=91.2$~GeV.  According to
the renormalization scheme~\cite{CWZ} we  used in our FO
calculation~\cite{zbb4}, this implies that the strong coupling constant runs
with $(n_f -1)$ flavours, where $n_f$ is the total number of flavours,
including the massive one.
For this reason, in order to compare the FO results with the resummed and the
truncated ones (where the the heavy flavour is treated as a light one), we
have to change the strong coupling constant, according to
\beq
  \asb^{(n_f-1)}(E) =  \asb^{(n_f)}(E) - \frac{2}{3}\, T_F \,\asb^2\,
\log\frac{E^2}{m^2}+{\cal O}\(\as^3\) \;,
\eeq
where we have used the renormalization group equation~(\ref{eq:RGE}) and the 
matching condition
\beq
\asb^{(n_f)}(m) = \asb^{(n_f-1)}(m) \;.
\eeq
This implies that the FO differential cross section, computed with
$\as^{(n_f-1)}$, 
\beq
   \left. \frac{d\s}{d\xp}\fix = a^{(0)} + a^{(1)}\, \asb^{(n_f-1)}(E)+
 a^{(2)} \, \asb^2(E)\;,
\eeq
acquires a contribution, proportional to the $a^{(1)}$ term, once written in
terms of $\as^{(n_f)}$
\beq
   \left. \frac{d\s}{d\xp}\fix = a^{(0)} + a^{(1)} \, \asb^{(n_f)}(E)+
 \lq a^{(2)} - \frac{2}{3}\, T_F \,  a^{(1)}\,
\log\frac{E^2}{m^2} \rq \asb^2(E)\;,
\eeq
where we have dropped all the arguments of the coefficients, for ease of
notation. 

\begin{figure}[htb]
\centerline{\epsfig{figure=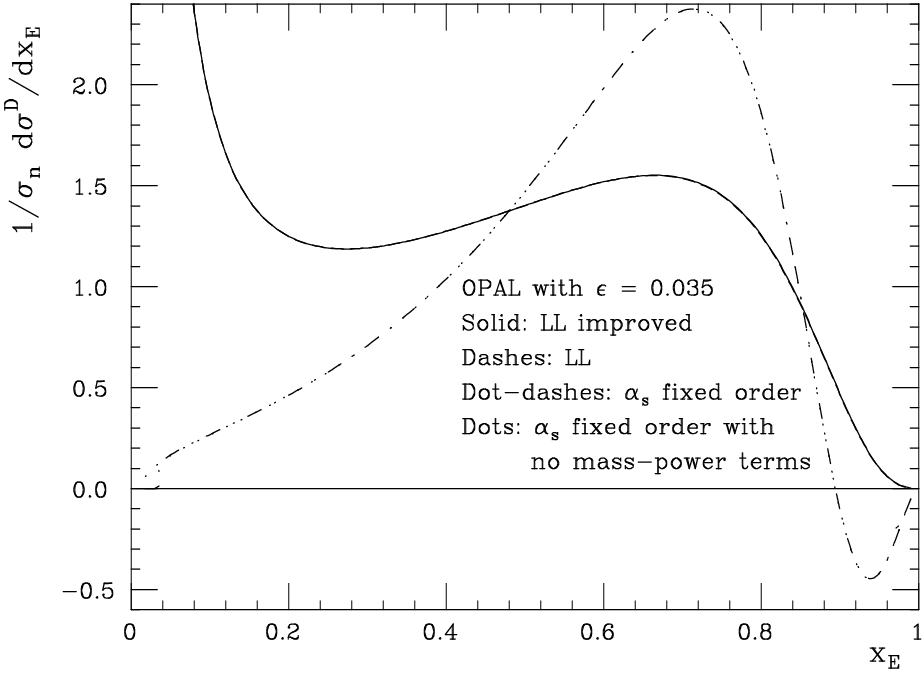,width=0.75\textwidth,clip=}}
\ccaption{}{ \label{fig:opal_as}
Fragmentation function $1/\s_{\rm tot} \, d\s^D/d\xe$ for
\OPAL.  The value of the Peterson $\ep$ parameter is fixed at
0.035. The dashed curve is almost hidden by the solid one.  }
\end{figure}

\begin{figure}[htb]
\centerline{\epsfig{figure=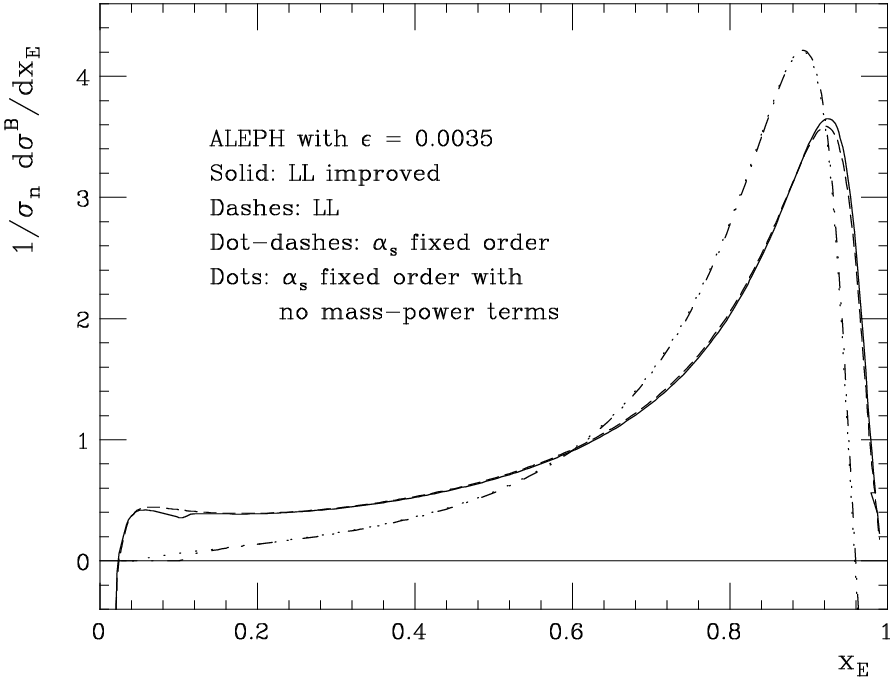,width=0.75\textwidth,clip=}}
\ccaption{}{ \label{fig:aleph_as}
Fragmentation function $1/\s_{\rm tot} \, d\s^B/d\xe$ for
\ALEPH.  The value of the Peterson $\ep$ parameter is fixed at
0.0035.  }
\end{figure}

In Figs.~\ref{fig:argus_as},~\ref{fig:opal_as}
and~\ref{fig:aleph_as}, we plot the following quantities:
\begin{figure}[htb]
\centerline{\epsfig{figure=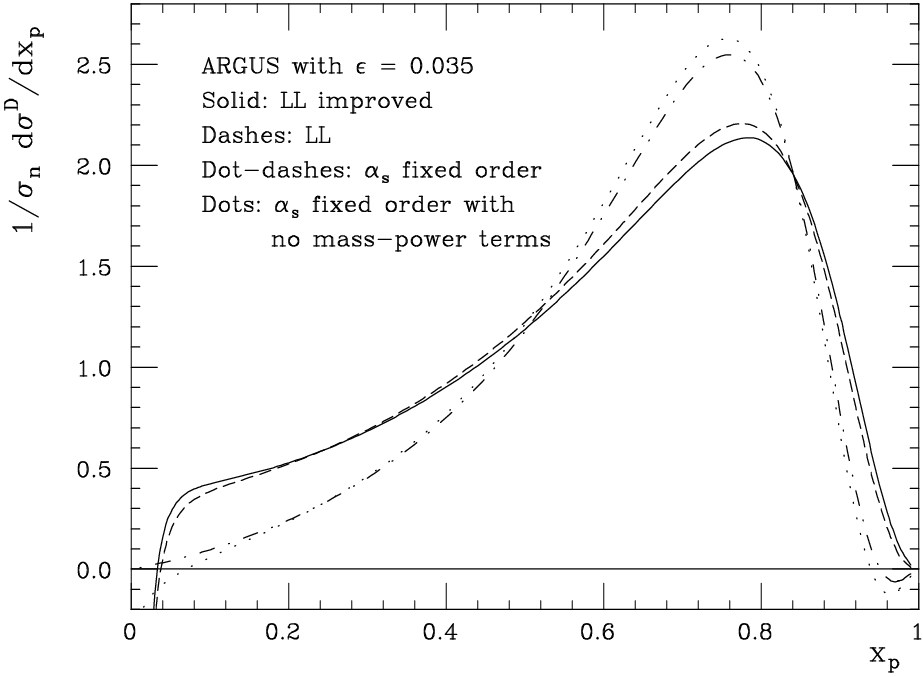,width=0.75\textwidth,clip=}}
\ccaption{}{ \label{fig:argus_as}
Fragmentation function $1/\s_{\rm tot} \, d\s^D/d\xp$ for
\ARGUS.  The value of the Peterson $\ep$ parameter is fixed at
0.035.  }
\end{figure}
\begin{itemize}
  \item[-] the dot-dashed line is the differential
  cross section at order $\as$.
  It corresponds to the sum of the terms up to the order
  $\as$ in Eq.~(\ref{eq:sig_fix_norm});
\item[-] the dotted line represents the cross section at order $\as$ 
  without power-suppressed mass effects.
  We see that such terms give a noticeable contribution
  only at \ARGUS, while they are completely negligible
  at \OPAL\ and \ALEPH;
\item[-] the dashed line represents the LL resummed cross
  section of Eq.~(\ref{eq:LL}).
\item[-] the solid line is the improved differential cross section,
  obtained by merging the LL resummed and the $\as$-order massive one.
  At \OPAL\ and \ALEPH\, the LL and the improved LL curves are almost
  identical.
  From this we infer that $\ord{\as}$ mass terms are small at \ARGUS\ energies,
  and completely negligible both for $b$ and for $c$ at LEP energies.
\end{itemize}

Notice that for very large and very small $x$, the distributions
we computed may become negative. This is an indication of the failure
of the perturbative expansion, due to the presence of large terms
proportional to powers of $\log(x)$ and $\log(1-x)$. These terms have not
been resummed in our approach.
We will discuss in more detail this problem in the next section.

\begin{figure}[htb]
\centerline{\epsfig{figure=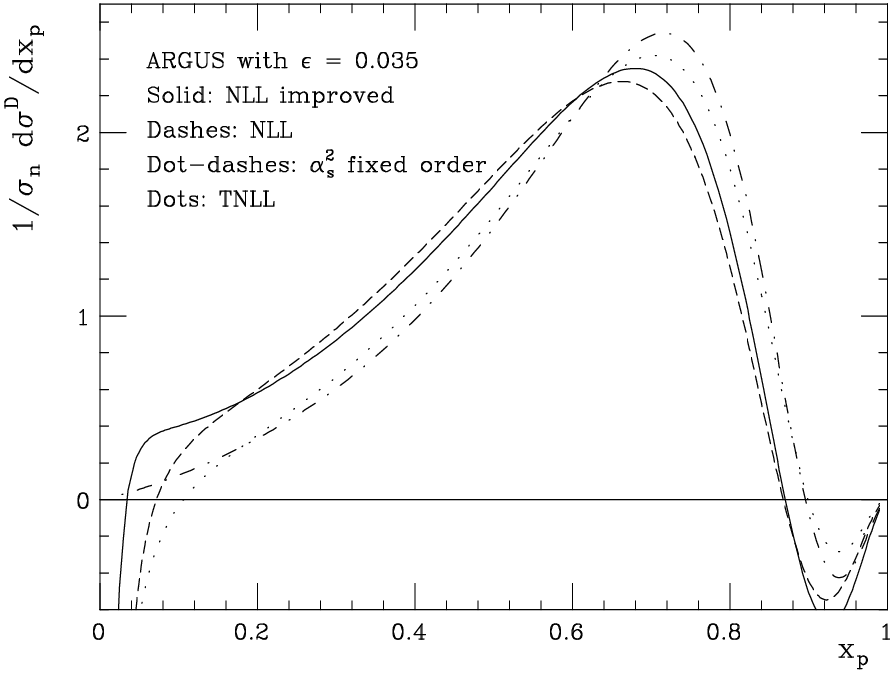,width=0.75\textwidth,clip=}}
\ccaption{}{ \label{fig:argus_as2}
Fragmentation function $1/\s_{\rm tot} \, d\s^D/d\xp$ for
\ARGUS.  The value of the Peterson $\ep$ parameter is fixed at
0.035.  }
\end{figure}

\begin{figure}[htb]
\centerline{\epsfig{figure=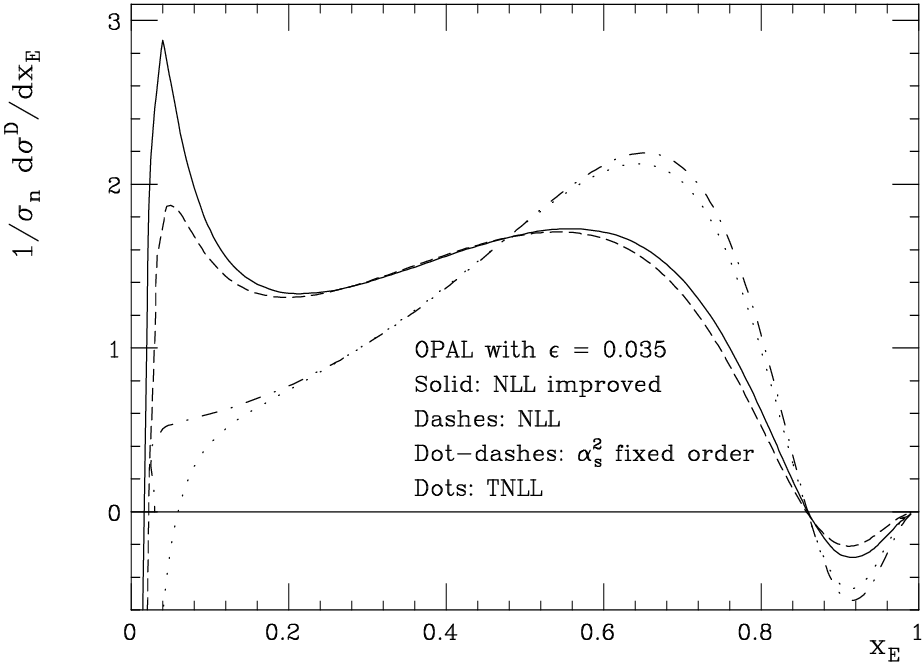,width=0.75\textwidth,clip=}}
\ccaption{}{ \label{fig:opal_as2}
Fragmentation function $1/\s_{\rm tot} \, d\s^D/d\xe$ for
\OPAL.  The value of the Peterson $\ep$ parameter is fixed at
0.035.  }
\end{figure}

\begin{figure}[htb]
\centerline{\epsfig{figure=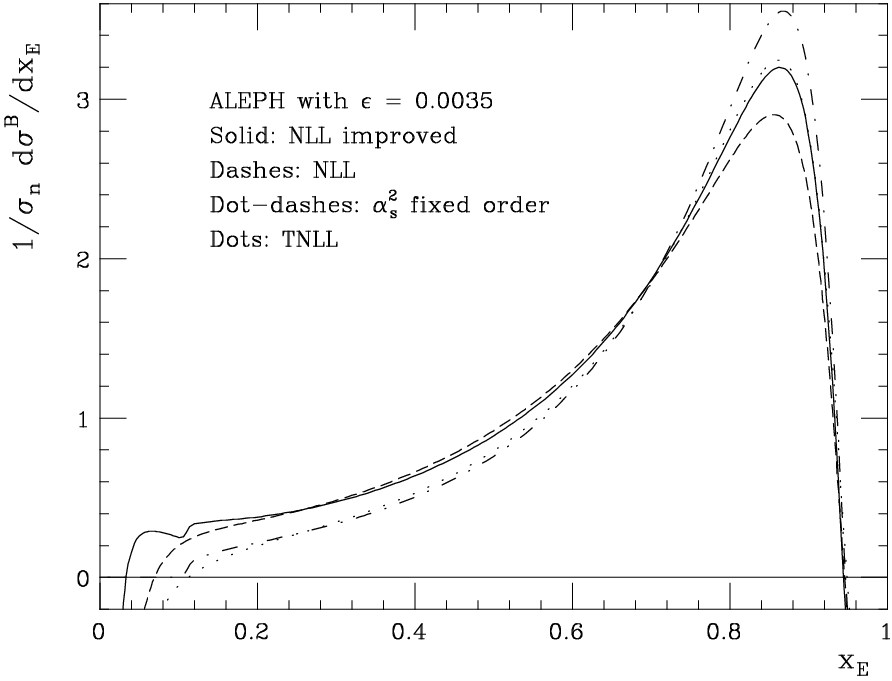,width=0.75\textwidth,clip=}}
\ccaption{}{ \label{fig:aleph_as2}
Fragmentation function $1/\s_{\rm tot} \, d\s^B/d\xe$ for
\ALEPH.  The value of the Peterson $\ep$ parameter is fixed at
0.0035.  }
\end{figure}

Figures~\ref{fig:argus_as2},~\ref{fig:opal_as2}
and~\ref{fig:aleph_as2} are similar to Figs.~\ref{fig:argus_as},
\ref{fig:opal_as} and~\ref{fig:aleph_as},
but they include next-to-leading effects.

Thus:
\begin{itemize}
  \item[-] the dot-dashed line is the differential
  cross section of Eq.~(\ref{eq:sig_fix_norm}) at order $\as^2$.
  All mass effects and the exact $\as^2$ term are present;
\item[-] the dotted line represents the TNLL cross section
  of Eq.~(\ref{eq:TNLL}). It differs from the previous curve for the absence
  of  mass power-suppressed effects and of NNLL $\as^2$
  contributions 
  (the terms not accompanied by large logarithms).
  We can see that, for the \OPAL\ and \ALEPH\ curves, where
  $m^2/E^2$ terms are much smaller than the \ARGUS\ ones, the effect of
  the NNLL $\as^2$ term is quite small, except for the small-$x$ region in
  the \OPAL\ cross section, where the 
  splitting mechanism of a gluon, coming from a primary light-quark pair,
  gives a sizable contribution; 
\item[-] the dashed line is the NLL resummed cross section
  of Eq.~(\ref{eq:sigNLLH}), normalized to $\s_{\rm n}$ of
  Eq.~(\ref{eq:sig_norm});
\item[-] the solid line is the improved cross section of
  Eq.~(\ref{eq:improvedH}), that takes account of the NLL
  resummed effects and of the mass and NNLL $\as^2$ terms.
\end{itemize}
We notice that, for the next-to-leading curves, the resummed cross
section tends to be softer than the fixed-order one, so that it will
need a harder non-perturbative fragmentation function (corresponding
to smaller values of $\ep$) in order to fit the data.  Furthermore, this effect
is much more pronounced at LEP energies, as one can expect.

Mass effects seem to be very small in this context. In general, we see that
they harden the fragmentation function. We thus expect that they will lead
to larger values of $\ep$ when fitting the data.

\section{Fit to the experimental data}
\label{sec:Fits}

We present now some fits to the experimental data in order to extract the
non-perturbative part of the fragmentation  functions.

We consider the following data sets
\begin{itemize}
 \item[-] $D^{*+}$ meson at \ARGUS~\cite{ARGUS}, at $E=10.6$~GeV,
 \item[-] $D^{*}$ meson at \OPAL~\cite{OPAL}, at $E=91.2$~GeV, 
 \item[-] $B$ meson at \ALEPH~\cite{ALEPH}, at $E=91.2$~GeV. 
\end{itemize}
Besides the LL and NLL fits (similar to that ones made in
Ref.~\cite{CacGre}), we present new fits with our improved cross
section.
Furthermore, we present fits where the initial conditions~(\ref{eq:ini})
for the fragmentation functions are taken in an exponentiated form (called
``NLL expon''). In $N$-space, the exponentiated initial conditions read
\beqn
\label{eq:ini_exp_N}
\Dh_Q(N,\mu_0,m) &=& \exp\lq \asb(\mu_0)\,
 d^{(1)}_Q(N,\mu_0,m) \rq \nn\\
\Dh_g(N,\mu_0,m) &=& \exp\lq \asb(\mu_0)\, d^{(1)}_g(N,\mu_0,m)\rq
-1\;.
\eeqn
These initial conditions are equivalent, from the point of view
of NLL resummation, to the ones in Eq.~(\ref{eq:ini}). They are introduced 
here
only to enhance higher order effects in the large-$x$ region.
In fact, these terms partially account for Sudakov behaviors in the
endpoint region~\cite{MeleNason,DokKhoze}.
We will see that their effect is quite dramatic, and, in particular,
that they remedy to the problem of negative cross sections at large $x$.
They are shown here just for the purpose of illustrating how
higher order perturbative terms may get rid of this problem.

The negative values in the small-$x$ region are related to the multiplicity
problem~\cite{BCM}. We will not try to remedy to them in this context,
since they are not very important in the experimental configurations
we consider.

We have fitted the data by $\chi^2$ minimization. With this procedure we
have fitted both the value of $\ep$ and the normalization, which was allowed
to float. We have kept  $\Lambda_{\rm QCD}^{(5)}$ fixed to $200$~MeV.

\begin{figure}[htb]
\centerline{\epsfig{figure=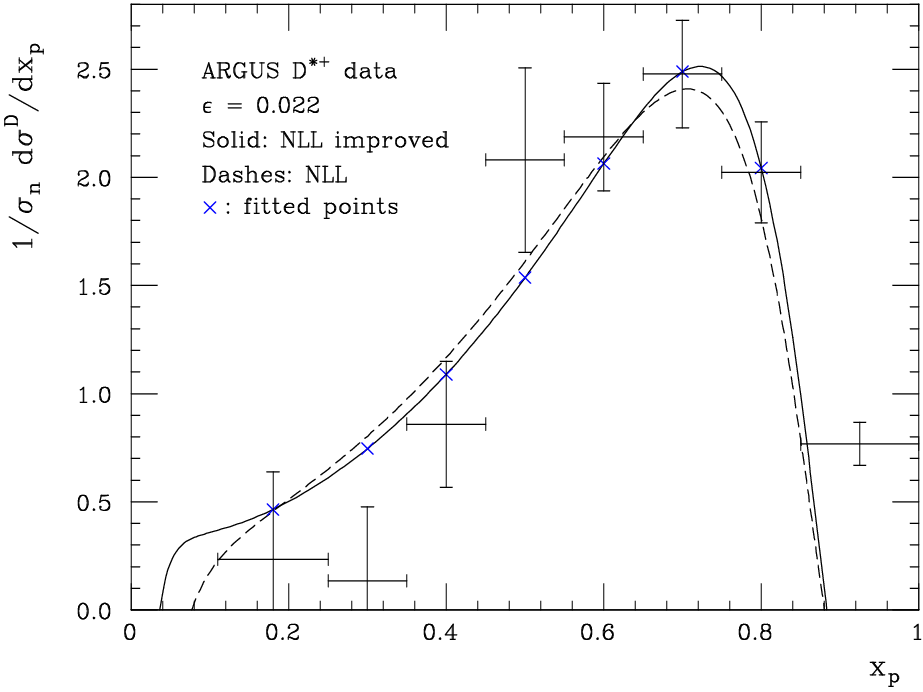,width=0.75\textwidth,clip=}}
\ccaption{}{ \label{fig:argus_nllimprov}
Best fit for the improved fragmentation function 
at \ARGUS.  In dashed line, the NLL
fragmentation function at the same value of $\ep$. }
\end{figure}

\begin{figure}[htb]
\centerline{\epsfig{figure=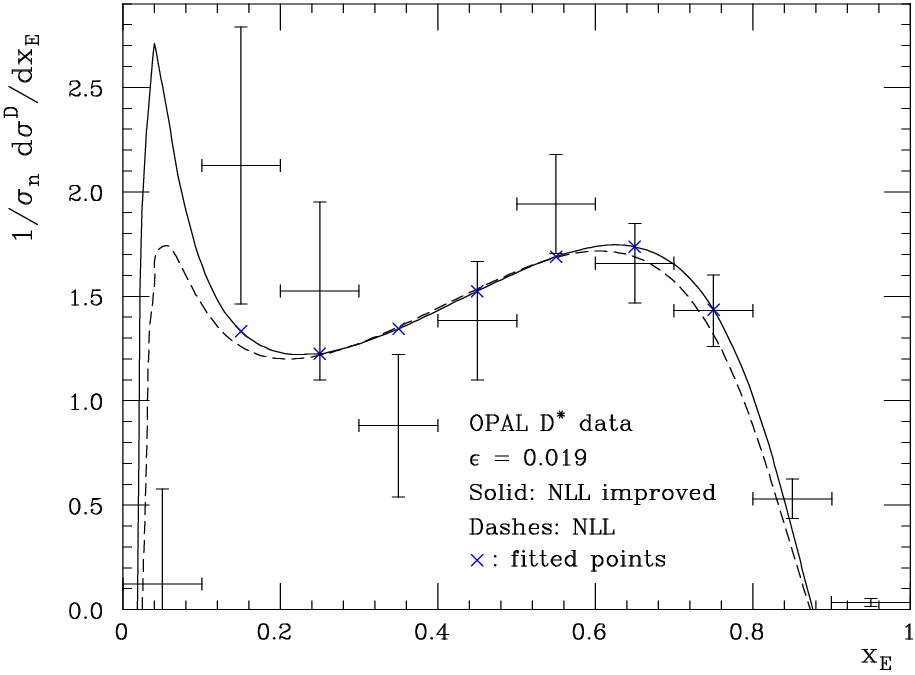,width=0.75\textwidth,clip=}}
\ccaption{}{ \label{fig:OPAL_nllimprov}
Best fit for the improved fragmentation function 
at \OPAL.  In dashed line, the NLL
fragmentation function at the same value of $\ep$. }
\end{figure}

\begin{figure}[htb]
\centerline{\epsfig{figure=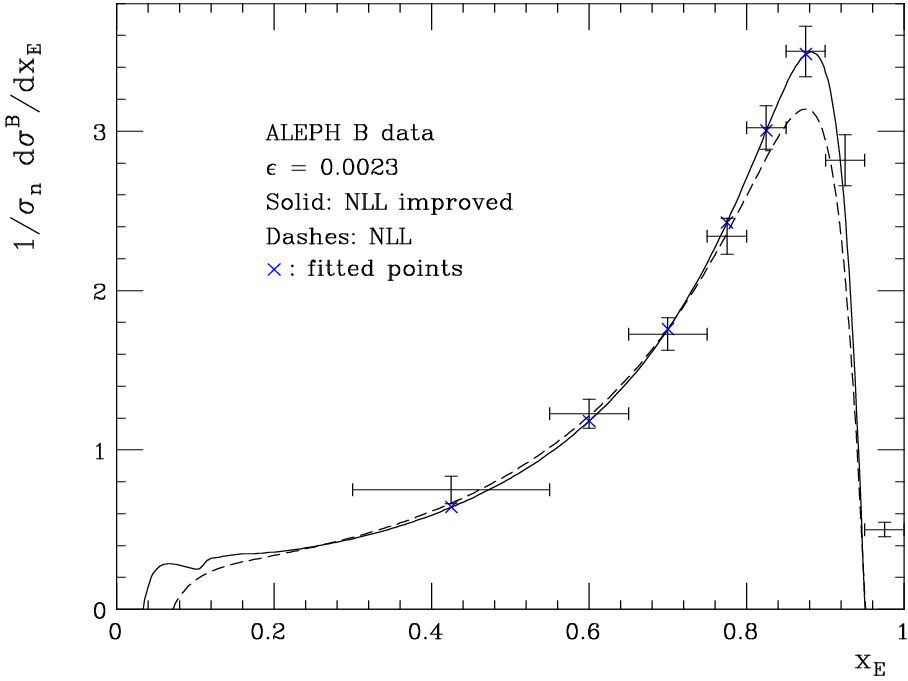,width=0.75\textwidth,clip=}}
\ccaption{}{\label{fig:aleph_nllimprov}
Best fit for the improved fragmentation function 
at \ALEPH.  In dashed line, the NLL
fragmentation function at the same value of $\ep$. }
\end{figure}

The results of the fits are displayed in Tables~\ref{tab:ep_valuesI} 
and~\ref{tab:ep_valuesII}.
\begin{table}[htbp]
 \begin{center}
 \leavevmode
\begin{tabular}{c||c|c|c}
 \hline\hline
$\ep\;(\chi^2/{\rm dof})$
& $\as$ fixed order & LL & LL improved \\
 \hline\hline
ARGUS $D$ & 0.058 (0.852) & 0.053 (2.033) & 0.054 (2.194) \\
OPAL $D$& 0.078 $\!\!\mbox{}^{(*)}\!\!$ (0.706) & 0.048 (1.008)  
& 0.048 (1.008) \\
ALEPH $B$& 0.0069 (4.607)  & 0.0061 (0.137)  & 0.0064 (0.137) \\
      \hline\hline
    \end{tabular}
    \caption{Results of the fit of the non-perturbative $\ep$
 parameter for the Peterson fragmentation function. The value of
 $\chi^2$/dof is given in parenthesis. The range of the fit is
 indicated in
Figs.~\ref{fig:argus_nllimprov}--\ref{fig:opal_nllexp} with small crosses.
$\mbox{}^{(*)}\!\!$ We have excluded the first three experimental  points for
 the $\as$ fixed-order fit to the \OPAL\ data.}
    \label{tab:ep_valuesI}
  \end{center}
\end{table}
\begin{table}[htbp]
 \begin{center}
 \leavevmode
\begin{tabular}{c||c|c|c|c}
 \hline\hline
$\ep\;(\chi^2/{\rm dof})$ &
$\as^2$ fixed order  & NLL  & NLL improved & NLL expon\\
 \hline\hline
ARGUS $D$& 0.035 (0.855)& 0.018 (1.234) & 0.022 (1.210)&
 0.0032 (1.493)\\
OPAL $D$& 0.040 $\!\!\mbox{}^{(*)}\!\!$ (0.769) & 0.016 (1.122) 
& 0.019 (1.066) & 0.0042 (1.152) \\
ALEPH $B$& 0.0033 (2.756) & 0.0016 (0.441)  & 0.0023 (0.635) & 
0.0003 (5.185) \\
      \hline\hline
    \end{tabular}
    \caption{Results of the fit of the non-perturbative $\ep$
 parameter for the Peterson fragmentation function. The value of
 $\chi^2$/dof is given in parenthesis. The range of the fit is
 indicated in Figs.~\ref{fig:argus_nllimprov}--\ref{fig:opal_nllexp} with
 small crosses. 
 $\mbox{}^{(*)}\!\!$ We have excluded the first three experimental  points for
 the $\as^2$ fixed-order fit to the \OPAL\ data.}
  \label{tab:ep_valuesII}
  \end{center}
\end{table}
The corresponding curves, together with the data, are shown
in Figs.~\ref{fig:argus_nllimprov}--\ref{fig:aleph_nllimprov}.


The full improved resummed result of Eq.~(\ref{eq:improvedH}) has been used
here.
For comparison, we have also plotted the NLL curves
computed at the same value of $\ep$.

\begin{figure}[htb]
\centerline{\epsfig{figure=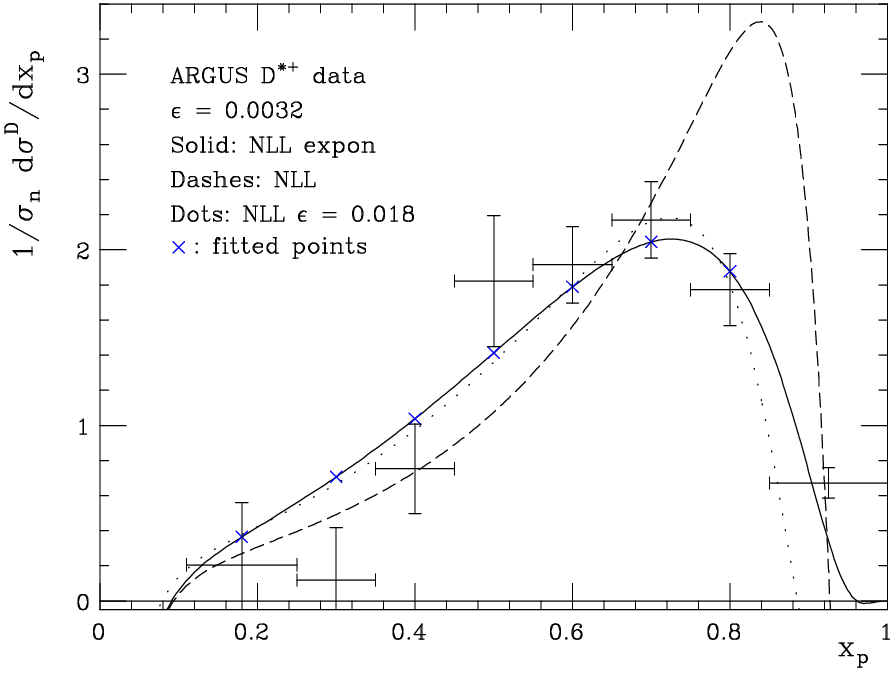,width=0.75\textwidth,clip=}}
\ccaption{}{ \label{fig:argus_nllexp}
Best fit for the NLL fragmentation function with exponentiated
starting condition
at \ARGUS.  In dashed line, the NLL
fragmentation function at the same value of $\ep$. In dotted line the NLL
best fit.}
\centerline{\epsfig{figure=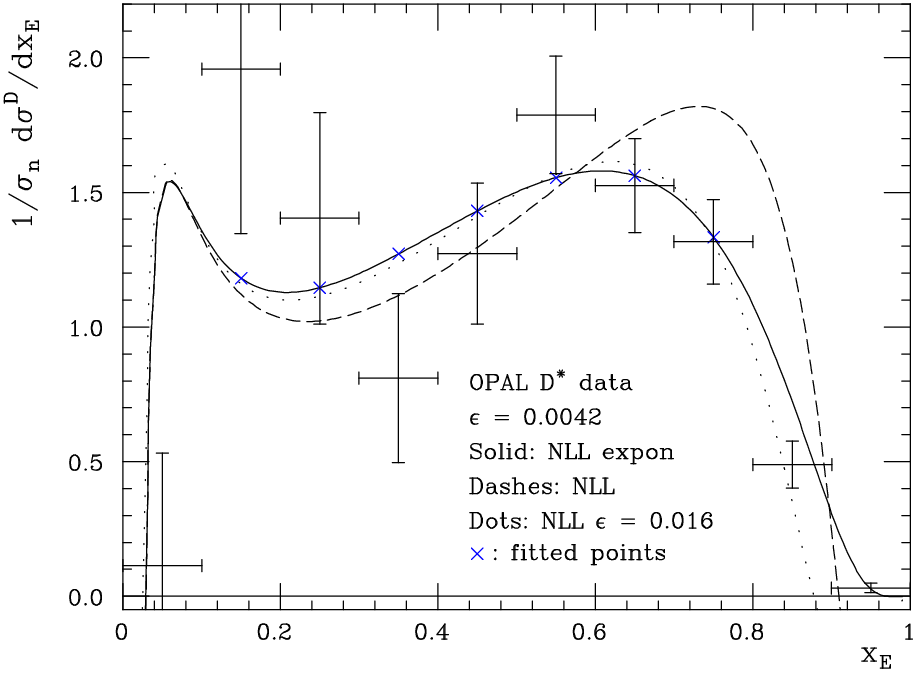,width=0.75\textwidth,clip=}}
\ccaption{}{ \label{fig:opal_nllexp}
Best fit for the NLL fragmentation function with exponentiated
starting condition
at \OPAL.  In dashed line, the NLL
fragmentation function at the same value of $\ep$. In dotted line the NLL
best fit.}
\end{figure}

In Figs.~\ref{fig:argus_nllexp} and~\ref{fig:opal_nllexp}, we 
illustrate the NLL differential cross section, obtained with the
exponentiated condition~(\ref{eq:ini_exp_N}).
We see a better behavior at $x\,\to\,1$, while the values of the $\ep$
parameter for the best fit are quite small.
For comparison, we have also plotted the NLL curves at the same $\ep$
value and the NLL best fit with the value of
$\ep$ taken from Tab.~\ref{tab:ep_valuesII}.

The small differences we find in the LL and NLL sectors, with respect to
the results of Ref.~\cite{CacGre},
are due to different range, normalization and
adjustment of physical parameters.

From the value of the $\ep$ parameter for \ARGUS, \OPAL\ and \ALEPH,
we can easily see that it scales nearly quadratically
in the heavy-quark mass, as expected.

\section{Conclusions}
\label{sec:conclusions}

In this work we have considered the heavy-flavour fragmentation
functions in $\epem$ annihilation.
We have devised and implemented a method by which all perturbative effects
that have been calculated so far can be included in the
computation of the fragmentation function. These include leading and
next-to-leading logarithmic resummation, fixed-order effects up to
$\ord{\as^2}$, and mass effects to the same order.

Our finding can be easily summarized as follows.  We generally find little
difference between our results and the NLL resummed calculation. This
indicates that mass effects are of limited importance in
fragmentation-function physics in $\epem$ annihilation.  On the other
hand, our calculation confirms the fact that, when NLL effects are
included, the importance of a non-perturbative initial condition is
reduced. For example, at \ARGUS\ energies, we see a strong reduction of
the $\ep$ parameter, from $0.053$ to $0.018$.  This reduction
is also 
observed in the fixed-order calculation, where $\ep$ goes from $0.058$
to $0.035$ when the ${\cal O}(\as^2)$ effects are included.



\clearpage

\relax
\def\pl#1#2#3{{\it Phys.\ Lett.\ }{\bf #1}\ (19#2)\ #3}
\def\zp#1#2#3{{\it Z.\ Phys.\ }{\bf #1}\ (19#2)\ #3}
\def\prl#1#2#3{{\it Phys.\ Rev.\ Lett.\ }{\bf #1}\ (19#2)\ #3}
\def\rmp#1#2#3{{\it Rev.\ Mod.\ Phys.\ }{\bf#1}\ (19#2)\ #3}
\def\prep#1#2#3{{\it Phys.\ Rep.\ }{\bf #1}\ (19#2)\ #3}
\def\pr#1#2#3{{\it Phys.\ Rev.\ }{\bf #1}\ (19#2)\ #3}
\def\np#1#2#3{{\it Nucl.\ Phys.\ }{\bf #1}\ (19#2)\ #3}
\def\sjnp#1#2#3{{\it Sov.\ J.\ Nucl.\ Phys.\ }{\bf #1}\ (19#2)\ #3}
\def\app#1#2#3{{\it Acta Phys.\ Polon.\ }{\bf #1}\ (19#2)\ #3}
\def\jmp#1#2#3{{\it J.\ Math.\ Phys.\ }{\bf #1}\ (19#2)\ #3}
\def\nc#1#2#3{{\it Nuovo Cim.\ }{\bf #1}\ (19#2)\ #3}
\def\jhep#1#2#3{{\it J.\ High Energy Phys.\ }{\bf #1}\ (19#2)\ #3}
\relax


\begin{thebibliography}{99}
\bibitem{MeleNason}
B.~Mele and P.~Nason, \np{B361}{91}{626}.
\bibitem{ColangeloNason}
G.~Colangelo and P.~Nason, \pl{B285}{92}{167}.
\bibitem{CacGre}
M.~Cacciari and M.~Greco, \pr{D55}{97}{7134}.
\bibitem{RandallRius}
L.~Randall and N.~Rius, \np{B441}{95}{167}.
\bibitem{Peterson}
C.~Peterson, D.~Schlatter, I.~Schmitt and P.~M.~Zerwas, \pr{D27}{83}{105}.
\bibitem{ALEPH}
D.~Buskulic et al., ALEPH Collaboration, \pl{B357}{95}{699}.
\bibitem{OPAL_b}
G.~Alexander et al., OPAL Collaboration, \pl{B364}{95}{93}.
\bibitem{SLD}
K.~Abe et al., SLD Collaboration, \pr{D56}{97}{5310}.
\bibitem{cgn98}
M.~Cacciari, M.~Greco and P.~Nason, \jhep{05}{98}{007}.
\bibitem{Braaten}
E.~Braaten, K.~Cheung, S.~Fleming and Tzu Chiang Yuan,
\pr{D51}{95}{4819}, \hepph{9409316};\newline
R.~L.~Jaffe and L.~Randall, \np{B412}{94}{79}, \hepph{9306201}.
\bibitem{zbb4}
P.~Nason and C.~Oleari, \np{B521}{98}{237};\newline
C.~Oleari, Ph.~D.~Thesis, \hepph{9802431}.
\bibitem{Rodrigo}
  G. Rodrigo, {\it Nucl. Phys. Proc. Suppl.} {\bf 54A}\ (1997)\ 60; \newline
  G. Rodrigo, Ph. D. Thesis,
  Univ. of Val\`encia, 1996, \hepph{9703359};\newline
  G.~Rodrigo, A.~Santamaria and M.~Bilenkii, \prl{79}{97}{193}.
\bibitem{Bernreuther}
  W. Bernreuther, A. Brandenburg and P. Uwer, \prl{79}{97}{189};
 A. Brandenburg and P. Uwer, \np{B515}{98}{279}.
\bibitem{frag97}
P.~Nason and C.~Oleari, \pl{B418}{98}{199}.
\bibitem{frag98-lett}
P.~Nason and C.~Oleari, \pl{B447}{99}{327}.
\bibitem{NasonWebber}
P.~Nason and B.~R.~Webber, \np{B421}{94}{473}.
\bibitem{Reinders}
L.~J.~Reinders, H.~Rubinstein and S.~Yazaki, \prep{127}{85}{1}.
\bibitem{CWZ} J.~Collins, F.~Wilczek and A.~Zee, \pr{D18}{78}{242}.
\bibitem{ARGUS}
H.~Albrecht et al., ARGUS Collaboration, \zp{C52}{91}{353}.
\bibitem{OPAL}
R.~Akers et al., OPAL Collaboration, \zp{C67}{95}{27}.
\bibitem{DokKhoze}
Yu.L.~Dokshitzer, V.A.~Khoze and S.~I.~Troyan, Workshop on jet studies at LEP
and HERA, DTP-91/04.
\bibitem{BCM}
A.~Bassetto, M.~Ciafaloni and G.~Marchesini, \pr{100}{83}{201}.
\end{thebibliography}
\end{document}